\documentclass[prl,aps,twocolumn,showpacs]{revtex4}

\usepackage{amsmath,graphicx       }
\begin{document}

\def\pd#1#2{\frac{\partial #1}{\partial #2}}

\title{\bf Turbulence in exciton--polariton condensates}
\author {Natalia G. Berloff}
\affiliation {Department of Applied Mathematics and Theoretical Physics,
University of Cambridge,  Cambridge, CB3 0WA\\
}
\date{22 October 2010}

\begin {abstract} Nonequilibrium condensate systems such as exciton-polariton condensates are capable of supporting a spontaneous vortex nucleation. The spatial inhomogeneity of pumping field or/and disordered potential creates velocity flow fields  that may become unstable to vortex formation. This letter considers ways in which  turbulent states of interacting vortices can be created. It is shown that by combining just two pumping  intensities it is possible to create a superfluid turbulence state of well-separated vortices, a strong turbulence state of de-structured vortices, or a weak turbulence state in which all coherence of the field is lost and motion is driven by weakly interacting dispersive waves. The decay of turbulence can be obtained by replacing an inhomogeneous pumping by a uniform one. We show that both in quasi-equilibrium and during the turbulence decay there exists an inertial range dominated by four-wave interactions of acoustic waves.
\end{abstract}
\maketitle
{\it Introduction.} The phenomenon of turbulence -- chaotic motion of vortices of many
different length scales -- is ubiquitous in nature, and   quantitative
understanding of it is a  notoriously difficult problem of classical physics.
Turbulence occurs in many usual fluid flows as well as in exotic systems such as
plasmas and superfluids. Vorticity 
in superfluids is quantized in units of $h/m$, where $m$ is the mass of the boson 
in contrast with continuously distributed vorticity of a classical 
Navier-Stokes fluid. In superfluids quantized vorticity is considered to be an evidence for a macroscopically occupied quantum
state that can be described by a classical complex-valued wave function $\psi({\bf x},t)$. Quantization of velocity circulation in superfluids leads to significant differences between superfluid turbulence (ST) and classical turbulence. On the other hand, at large Reynolds numbers  the motion of well--separated vortices in an incompressible classical  flow may have similar features to ST. In this case the vortex dynamics in superfluids is almost  classical in accordance with  the Biot-Savart law (BSL). The decay of the turbulence (loss of the vortex line density) occurs due to dissipative effects induced by interactions with a normal fluid component (with a thermal cloud).

Recently by introducing an external oscillatory perturbation in a trapped atomic BEC it became possible to  obtain  a disordered system of many topological defects~\cite{henn}. The dynamics of this matter field differs from  both  dynamics of vortices in classical turbulence and in superfluid helium turbulence. Firstly, the characteristic distance between vortices is comparable to their core sizes, so the chaotic behavior is seen on the level of a single vortex, secondly, these vortices are not structured, so they do not obey BSL, finally, 
the system is in a strongly non-equilibrium state. These creates a novel nontrivial regime of a classical complex matter field --- ``strong turbulence'' state -- whose evolution is quite different from that of ordered condensate. In analogy with other nonlinear systems such as plasmas, fluids and nonlinear optics, apart from the regime of strong turbulence there  exists the regime of weak turbulence where  all phases of the complex amplitudes of the matter field are random. Recently~\cite{chen} these three regimes (superfluid, strong turbulence and weak turbulence) have been observed at different temperatures in 2D cold atomic gases, showing a universal scaling. The weak turbulence  plays crucial role in kinetics of 
Bose-Einstein
condensation~\cite{berloff02}. It was shown that a strongly non-equilibrium Bose gas evolves from the regime  of weak
turbulence to superfluid turbulence,  via states  of strong
turbulence in the long-wavelength  region of  energy space.  An important question remains whether it is possible to force a condensate system to pass through these stages in a reverse order. It has been suggested ~\cite{berloffView} that if a sufficiently  strong external perturbation is applied to the trap, it is in principle possible to obtain  the weak turbulence state. When this is done it will lead to a  discovery of  nontrivial transitional regimes of classical matter fields in atomic systems~\cite{bognato10}. 

In the last few years the Bose-Einstein condensation has been achieved in solid state systems~\cite{BEC_solidstate}, such as microcavities, ferromagnetic insulators and within superfluid
phases of $^3$He.
Microcavity exciton-polaritons are quasi-particles that consist of superpositions
  of photons in semiconductor microcavities and excitons in quantum
  wells. The Bragg reflectors confining photon component are imperfect, so exciton-polariton have finite life time and and have to be continuously re-populated. Such combination of pumping and decay leads to quasi-particle flow even at steady states of the system. At sufficiently low densities these quasi-particles can form a Bose-Einstein condensate, so the  many particles quantum
system can be described by a
classical equation in a form of the
 complex Ginzburg-Landau equation (cGLE) \cite{keeling08, wouters07}:
\begin{equation}
  \label{GPE}
  2i\partial_t\psi
 =
  \left[
    - \nabla^2 
    + |\psi|^2
    + i \left(\alpha - i \eta \partial_t \psi - \sigma |\psi|^2 \right)
  \right]
  \psi,
\end{equation}
where  $\alpha$ is an effective gain that represents intensity of the pumping field, $\sigma$  represents nonlinear losses. The unit of length is a healing length $\xi=\hbar/\sqrt{2mU\rho_\infty}$ that defines the size of the vortex core and the unit of time is $\hbar/2U\rho_\infty$, where $U$ is the strength of a $\delta$-function interaction potential. We shall assume that $\alpha=\alpha_0$ is constant away from some localised nonuniformities and so the number density there is $\rho_\infty.$ It is possible to include a disorder potential of the microcavity by adding $V_{\rm ext}({\bf x})\psi$ to the right-hand side of Eq.~(\ref{GPE}).

This equation is  a mean-field description of the
condensate; it can also be derived from the saddle point
in a path integral formalism~\cite{szymanska07}. In the absence of pumping and dissipation Eq.(\ref{GPE}) reduces to the Gross-Pitaevskii equation  describing an equilibrium Bose-Einstein condensate. The energy relaxation has been noted to be of importance in experiments  on extended 1D waveguides~\cite{wertz10, liew10}. These effects can be included in the cGLE by means on a parameter $\eta$~\cite{wouters10:relax}. This is the same  term that has been incorporated into the Gross-Pitaevskii equation  to represent a dissipation of the condensate component due to interactions with a thermal cloud \cite{pitaevskii59}.
\begin{figure}[t!]
\caption{ (color online) The time snapshots of the density $|\psi|^2$  of the fields $\psi$ obtained by numerical integration of Eq. (\ref{GPE}) for $\eta=0, \sigma=0.3$ and (i)  $\alpha({\bf x})=2$ for $|{\bf x} - (\pm 5,0)| < 2$ and $1/2$ otherwise,  t=100 (left panel) and (ii)  $V_{\rm ext}=x^2+y^2$ and  $\alpha({\bf x})=5$ for $x^2+2y^2<64$ and $\alpha=-0.5$ otherwise (right panel). Red ellipse indicates the pumping spot. Luminosity of the density plots is proportional to density. Vortices are seen as black dots. }
 \centering
\bigskip
\includegraphics[height=1.5 in]{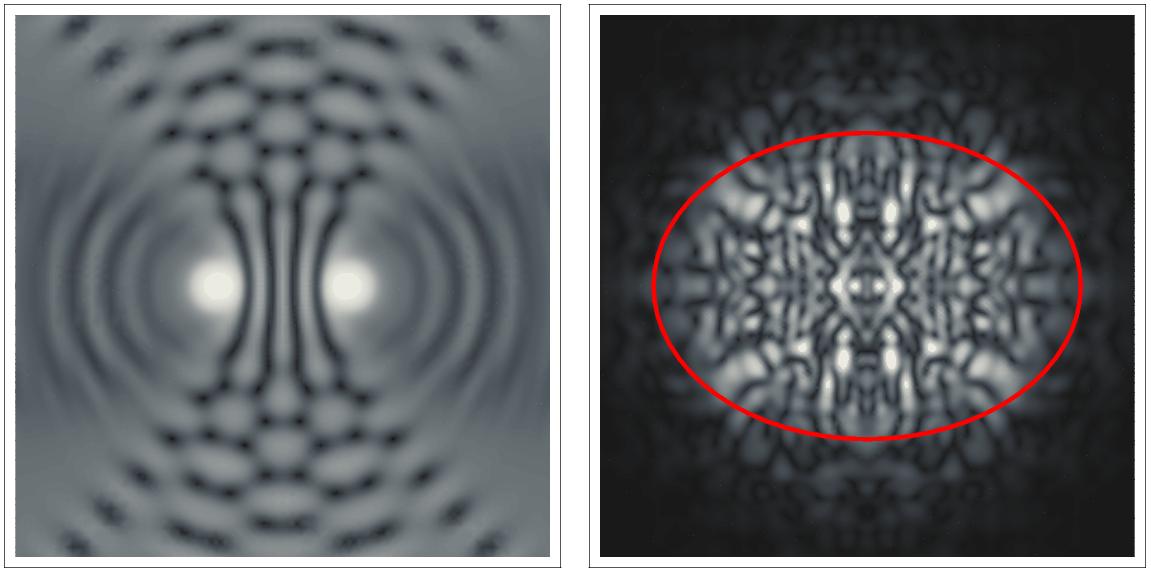}

\label{fig1}
\end{figure}
 The turbulence and mechanisms of the vortex generation in equilibrium condensates are well known. These include  (i) interactions of finite amplitude sound waves (e.g. energy exchange between  rarefaction pulses may lead to vortex formation)~\cite{berloff04}; (ii) existence of critical velocities of the flow (e.g. moving objects generate vortices if the Landau critical velocity is reached on their surfaces~\cite{movingbodies}); (iii) modulational instabilities of density variations (e.g. transverse instability of a dark soliton in 2D generates vortices)~\cite{dark}. Some of these mechanism may produce vortices in the cGLE as well. For instance, the flow of exciton-polaritons about a spatially extended defect may produce vortex pairs of opposite circulation depending on the flow velocity~\cite{ciuti10}.  In addition, Eq.~(\ref{GPE}) with gain and dissipation can form vortices by other physical mechanisms. For instance, an inhomogeneity of the  pumping or/and disorder potentials form steady currents which may produce vortices through pattern forming symmetry breaking mechanism~\cite{keeling08}. 

Although the formation of vortices has been observed in experiments~\cite{lagoudakis08} they seem to appear due to the  intrinsic disorder potential in
CdTe. The vortices become pinned at the local minimum of such potential and remain stationary. So the conditions in which a turbulent state of matter can be obtained in exciton-polariton condensates  remained unclear.  The purpose of this letter is to suggest how the turbulent state can be created in such a system, to study the properties and structure  of the turbulence, and to propose how different regimes can be detected experimentally. It will be shown that turbulence can be created by deliberately designed pumping fields, and depending on characteristics of such fields the system can reach various regimes of turbulence from superfluid turbulence to strong and finally weak-turbulent state.

{\it Vortex formation.} To illustrate the basic mechanism that drives the formation of vortices we first consider a pumping field in a form of a step function in 1D, so that $\alpha=\alpha_1+\alpha_0, \sigma=\sigma_1, \eta=\eta_1$ for $x<0$ and $\alpha=\alpha_0, \sigma=\sigma_0, \eta=\eta_0$ for $x>0$. The steady state  mass continuity and Bernoulli equations resulting from the Madelung transformation $\psi=\sqrt{\rho}\exp{i S}$ applied to Eq.~(\ref{GPE}) are $\mu=u^2+\rho -  d^2\sqrt{\rho}/2\sqrt{\rho}\,dx^2$ and $d(\rho u)/dx=(\alpha-\eta \mu - \sigma \rho)\rho$ where $\rho$ is the 
number density, $u=S'(x)$ is the velocity and  the chemical potential $\mu$ is introduced by $2 i \partial_t \psi = \mu \psi$. Away from large density fluctuations we can drop the quantum pressure term $d^2\sqrt{\rho}/2\sqrt{\rho}\,dx^2$. We expect that $ u\rightarrow0 $ as $x\rightarrow -\infty$, so $ \mu\rightarrow(\alpha_1+\alpha_0)/(\sigma_1+\eta_1) $. As $x\rightarrow \infty$, therefore, there will be a steady current $u=[\alpha_1(\eta_0+\sigma_0)+\alpha_0(\eta_0-\eta_1+\sigma_0-\sigma_1)/\sigma_0(\eta_1+\sigma_1)]^{-1/2}$ generated by the step. The presence of boundaries or other sources of outflow generate interference fringes seen, for instance, in recent experiments in 1D~\cite{wertz10}. In 2D the fringes that meet at nonzero angles evolve into a pair of vortices of opposite circulation as seen on the left panel of Fig.~\ref{fig1}. The mechanism leading to vortex formation in this case is analogous to the transverse instability of a density depletion in a conservative Gross-Pitaevskii equation~\cite{dark}: the motion of grey solitons is inversely proportional to their depth, so modulation in the transverse direction forces different parts of the front to move with different velocities leading to vortex pair formation. 
This suggests that the several sources of such flows  may continuously generate  a large number of vortices leading to a turbulent flow.  Another possibility to create a turbulent flow  is related to the formation of vortex lattice in a harmonic trapping potential due to an instability of a non-rotating solution~\cite{keeling08}. By removing the circular symmetry of either  the trapping potential or pumping field it is possible to create  a turbulent flow of vortices  instead of a regular vortex lattice (see the right panel of Fig.~\ref{fig1}).
\begin{figure}[t!]
\caption{ (color online) The evolution of the density of vortices as a function of time. For time $t<500$ the pumping is nonuniform as discussed in the text. At time $t=500$ the nonuniformity of the pumping field is removed and the vortex density decays linearly initially, as the inset shows. As the density of vortices decreases, the system reaches the superfluid turbulence regime of well-separated vortices with a logarithmic decay~\cite{nazarenko06}. These decay rates are in contrast with a power-decay rates of the order $t^{-3/4}$ in classical 2D viscous fluids and in the limit of the cGLE equation with zero dispersion~\cite{classvort}.  }
 \centering
\bigskip
\includegraphics[height=2 in]{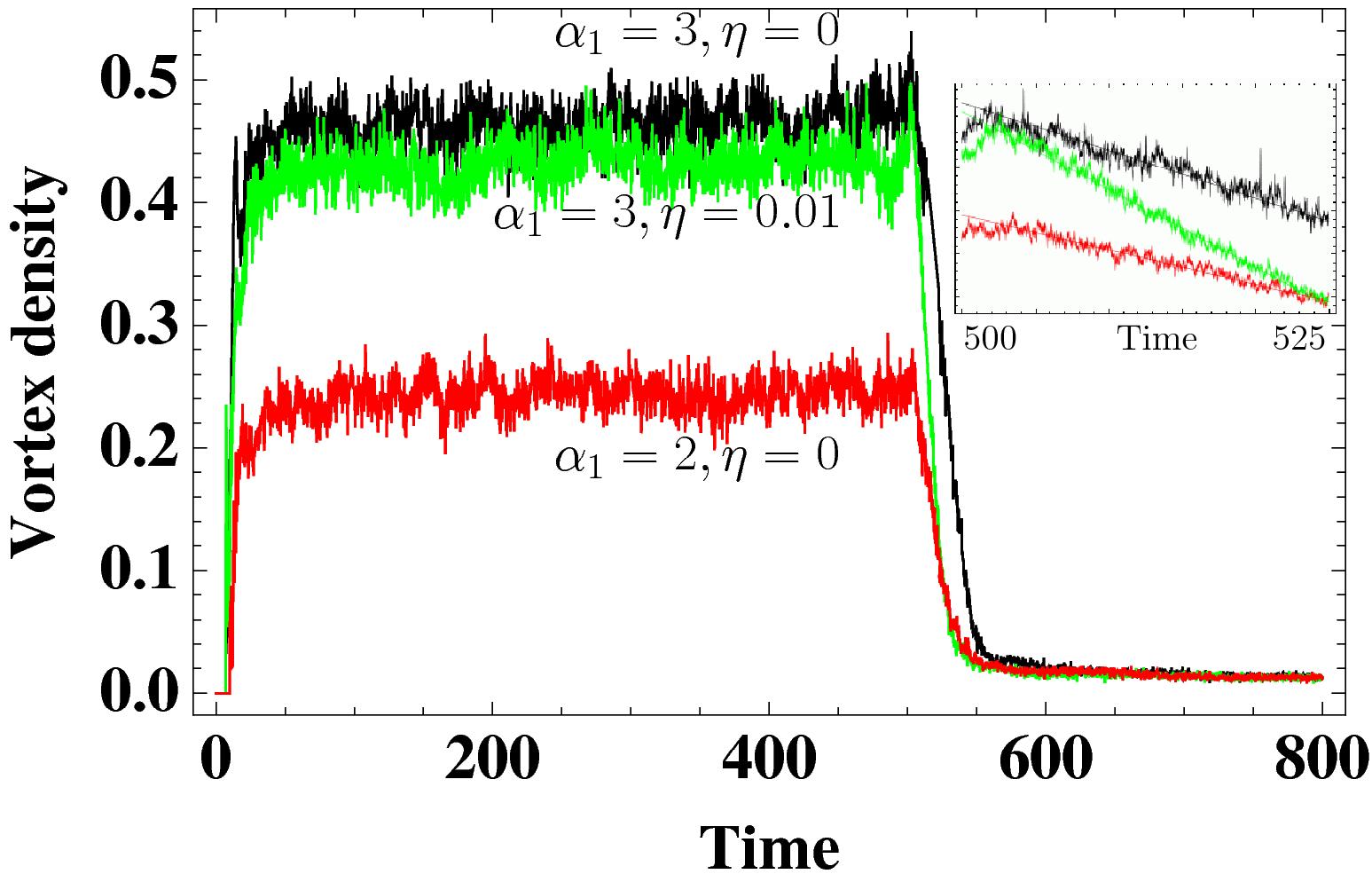}
\label{fig3}
\end{figure}

{\it Numerical set-up.} In order to engineer a turbulent formation and interaction of vortices we shall consider an inhomogeneous pump $\alpha({\bf x})$ that can be obtained by passing the laser beam through a spatial phase (light)  modulator. This will be even further simplified by assuming that only two laser intensities are allowed: the background with a superimposed set of almost periodical spots of a higher intensity, so that $\alpha({\bf x})=\alpha_0$ everywhere except for  ${\bf x}$ inside the circles $|{\bf x} - {\bf a}_i|^2 < c_i$ where $\alpha({\bf x}) =\alpha_1+\alpha_0$ and $c_i = c + \chi_i, {\bf a}_i=(a_1\pm i T+\delta_i,a_2\pm iT+\phi_i)$, $T$ is the period of spots, $c$ is the square of the spot radius, $i=0,1,2...$ and $\chi_i, \delta_i$ and $\phi_i$ are random displacements of the order of the healing length. Both $\eta$ and $\sigma$ take different values for different pumping intensities. In practice, setting different values for these quantities does not change the qualitative behaviour of the system. In what follows both $\eta$ and $\sigma$ with be set to be  constants across the fields~\cite{comment1}.

Through the time evolution  one can observe the formation of vortices until their number  begins to fluctuate about a constant value; see Fig.~\ref{fig3}. The larger  difference between the two pumping intensities, $\alpha_1$, leads to the faster outflows and a larger number of vortices generated. The relaxation has a negative effect on the number of vortices (compare the vortex densities for $\alpha_1=3$ and $\eta=0$ or $\eta=0.01$ on Fig.~\ref{fig3}). At time $t_s=500$ (well after the quasi-equilibrium is reached) we remove the nonuniformity of the pump by setting $\alpha_1=0$. After that the vortices start annihilating each other leading to the decay of the turbulence. This stage can be compared and contrasted with the wave turbulence of the Gross-Pitaevskii equation where the  dissipation is at a given (high)  momenta and so has a different physical meaning~\cite{nazarenko06}. 

\begin{figure}[t!]
\caption{ (color online) The  second moment of the correlation function $g_2$ as a function of $\alpha_1$. Starting from an initial constant density profile the nonuniform pumping is applied, so that $g_2$ rapidly grows reaching a quasi-stationary state after $t\sim 20$. After that $g_2$ fluctuates about a constant value. The blue dots ($\eta=0$) and red squares ($\eta=0.01$) show the average of the $g_2$ during the time interval $[20,t_s]$. The time snapshots of the normalized density $|\psi|^2$  of the fields $\psi$ are shown for  superfluid turbulence state (left inset), strong turbulence (bottom inset) and weak turbulence state (top inset) for $t < t_s.$}
 \centering
\bigskip
\includegraphics[height=2.2 in]{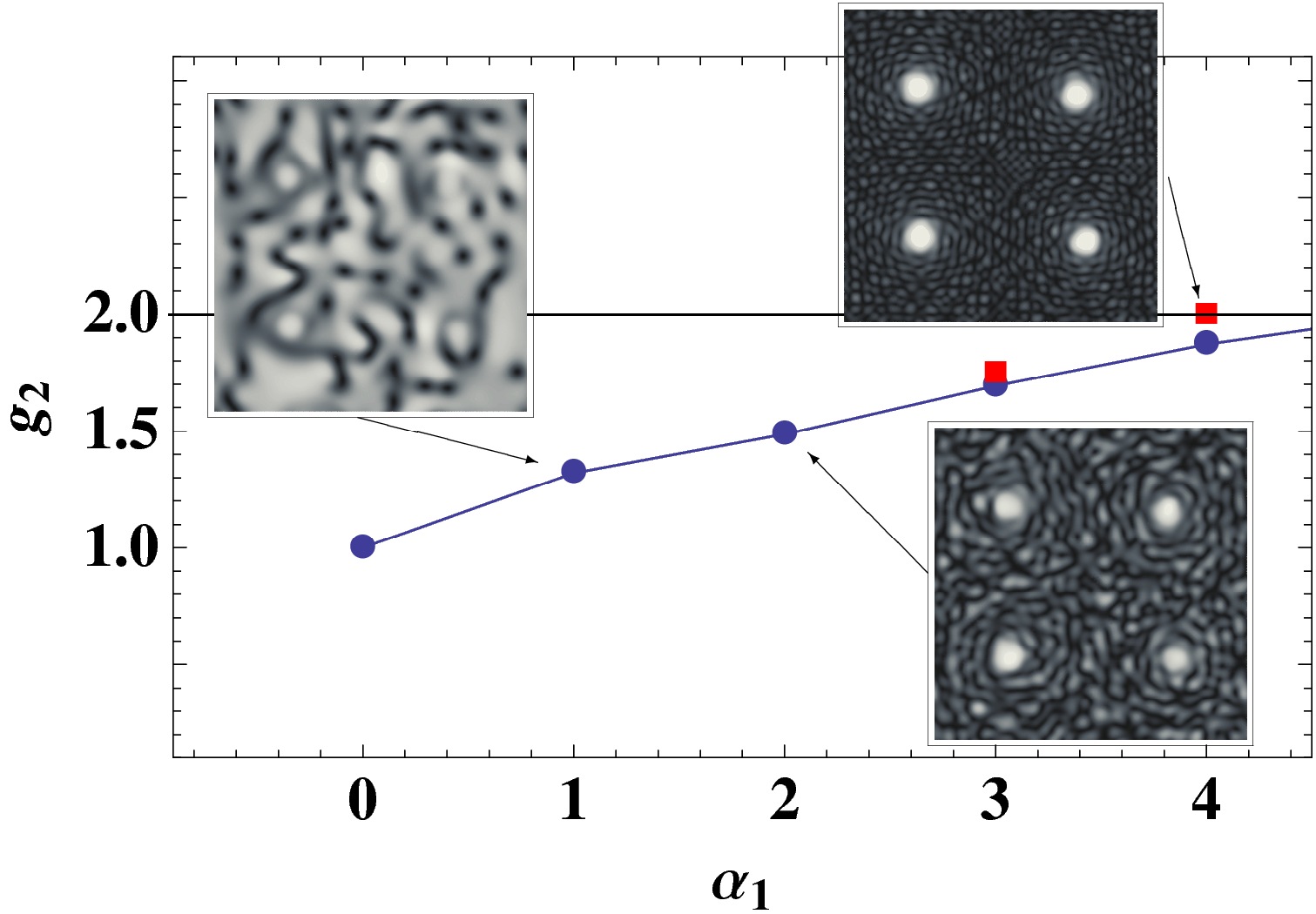}
\label{fig2}
\end{figure}

By tuning  the nonuniformity of the pumping field it is possible to reach different turbulent regimes. If the difference between intensities, $\alpha_1$, is below a threshold or the distance between the spots of higher intensity is large, no vortices will be created. In a case of a moderate $\alpha_1$ and only few spots  a set of several well-formed well-separated vortex pairs is created and the system is in a superfluid turbulence state (see the left panel of Fig.~\ref{fig1} and the left inset of  Fig.~\ref{fig2}). By increasing the    difference between intensities $\alpha_1$ it is possible to create the state of strong turbulence (where vortex cores start 
to overlap; see the top inset of Fig.~\ref{fig2}). It is, therefore, tempting to see if the system can be driven even further to enter the regime of weak turbulence in which all coherence is lost and all Fourier amplitudes have random phases. To verify this we calculated the second moment of the correlation function $g_2=\langle |\psi|^4\rangle/\langle |\psi|^2\rangle^2$. By Wick's theorem the state of the weak turbulence corresponds to $g_2=2$. As shown on Fig.~\ref{fig2} by raising $\alpha_1$ it is possible for the system to reach the weak turbulence state. Note that  the relaxation $\eta$ increases $g_2$. This occurs because the relaxation increases the rate at which vortex pairs annihilate by bringing the vortex cores closer to each other; this effect can be seen on Fig.~\ref{fig3} showing the number of vortices in quasi-equilibrium. The energy released from vortex annihilation becomes converted into acoustic energy therefore increasing $g_2$.

In order to describe the turbulence in the Eq.(\ref{GPE}) we shall assume 
that there exists an inertial range in the momentum space and that the role of 
pumping and dissipation is insignificant there. The evolution equation for the wave spectrum defined by $\left< a_{{\bf k}_1} a_{{\bf k}_2}^* \right> = n_{{\bf k}_1} \delta({\bf k}_1-{\bf k}_2)$, with $a_{\bf{k}}$ being 
the Fourier transform of $\psi$ and  ${\bf k}_i$ are discrete wave vectors, can be obtained by using a random phase approximation and expanding in small  nonlinearity~\cite{zakharov92}. The equation takes the form $
\partial_t n_{{\bf k}_1}(t) =  \int \!\!d^2k_2 d^2k_3 d^2k_4 W_{k_1,k_2;k_3,k_4} \times \ \left(n_{{\bf k}_3}n_{{\bf k}_4}n_{{\bf k}_1}\!+\! n_{{\bf k}_3}n_{{\bf k}_4} n_{{\bf k}_2}\!-\!
n_{{\bf k}_1}n_{{\bf k}_2}n_{{\bf k}_3}\!-\!n_{{\bf k}_1}n_{{\bf k}_2} n_{{\bf k}_4}\right)$, where $W_{k_1,k_2;k_3,k_4} = \frac{4\pi}{(2\pi)^{2}} 
\delta({\bf k}_1+{\bf k}_2-{\bf k}_3-{\bf k}_4)$ $ 
\delta(k_1^2+ k_2^2-k_3^2-k_4^2)$.
Two solutions of this evolution equation  correspond to a thermodynamic equipartition of the total kinetic energy $E=\int k^2 n_k\, d{\bf k}$, so that $n_k \sim k^{-2}$ and to an equipartition of the total number of particles $N=\int n_k\, d{\bf k}$, so that $n_k \sim {\rm const}$. These correspond to the two limits of the Rayleigh-Jeans distribution $T/(k^2+\mu)$, where $T$ is the temperature.

\begin{figure}[t!]
\caption{(color online) The wave spectrum $\log(n_k)$ vs $\log (k)$ for the state of strong turbulence established for parameters $\eta=0,  \sigma=0.3$ at $t=450$  (nonuniform pumping field with $\alpha_1=3, \alpha_0=1/2$, top blue curve)  and at $t=650$ (uniform pumping filed $\alpha_1=0, \alpha_0=1/2$, bottom red curve).  Lines corresponding to $n_k \sim const$ and $n_k \sim k^{-2}$ are included. Inset shows the wave spectrum for the state of superfluid turbulence $\alpha_1=1$ with $n_{\bf k} \sim k^{-2}$ spectrum of the inertial range and for the transitional state $\alpha_1=2$, both for $t < t_s$. }
 \centering
\bigskip
\includegraphics[height=2.2 in]{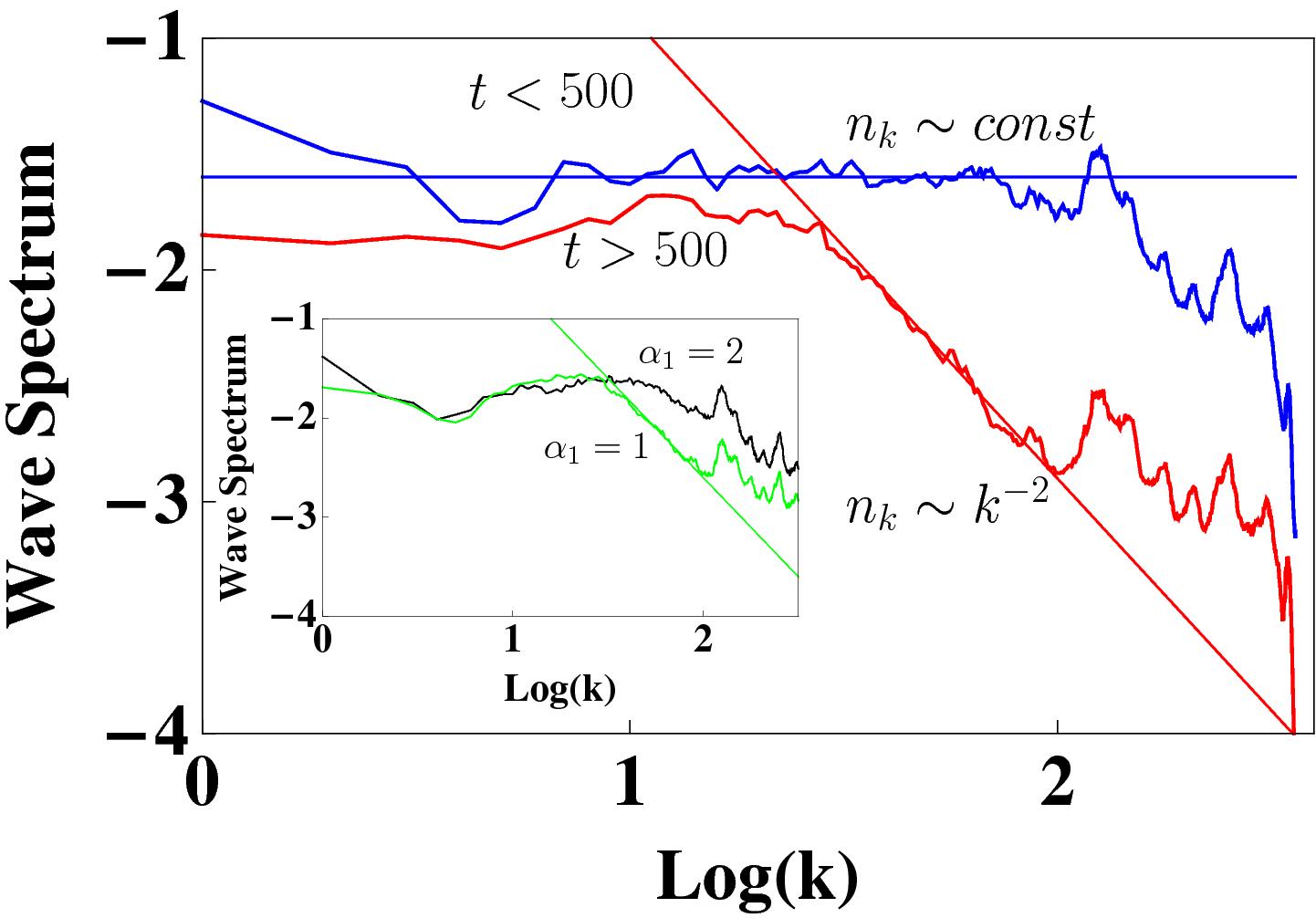}
\label{fig4}
\end{figure}

We verified the existence of the inertial range in our simulations. Although the system is in a quasi-equilibrium rather than in the true thermodynamical equilibrium we observed both spectra. For the nonuniform pumping the wave spectrum shows the particle equipartitions for strong turbulence (see the top (blue) curve of  Fig.~\ref{fig4}), whereas the superfluid turbulence spectra corresponds to energy equipartition (the Kolmogorov-Zakharov energy cascade $n_k \sim k^{-2}$); see the grey (green) curve on the inset of Fig.~\ref{fig4}. During the turbulence decay stage the wave spectrum corresponds to  $n_k \sim k^{-2}$; see the bottom (red) curve of Fig.~\ref{fig4}. This suggests that at these intermediate scales of the inertia range the turbulence is dominated by four--wave interactions and the wave field  is weakly nonlinear and dominated by acoustic modes.

In summary, we proposed a way  to generate various regimes of turbulence in nonequilibrium condensates, such as exciton--polariton condensates. By designing a nonuniform  pumping field that leads to sufficiently strong interacting fluxes it is possible to create the superfluid turbulence with well separated quantised vortices, the strong turbulence with overlapping and de-structured vortices or the weak turbulence state with a complete loss of coherence. The nonequilibrium condensates, therefore, are new and exciting systems with a nontrivial evolution of complex matter field with turbulence that may span regimes fundamentally different from the
classical fluid turbulence.

\acknowledgments{The author acknowledges useful discussions with A.~Amo, C.~Ciuti, J.~Keeling and  B.~Svistunov.}

\end{document}